\documentstyle[prl,aps]{revtex}
\begin{document}
\draft
\twocolumn[\hsize\textwidth\columnwidth\hsize\csname
@twocolumnfalse\endcsname

\title{Semiclassical wave functions and
energy levels of Bose-condensed gases\\
in spherically symmetric traps}

\author{Andr\'as Csord\'as$^1$, Robert Graham$^2$,
P\'eter Sz\'epfalusy$^{3}$}

\address{$^1$Research Group for Statistical Physics of the
Hungarian Academy of Sciences,\\
M\'uzeum krt. 6--8, H-1088 Budapest,
Hungary \\
$^2$Fachbereich Physik,
Universit\"at-Gesamthochschule Essen,
45117 Essen,Germany \\
$^3$Institute for Solid State
Physics, E\"otv\"os University,
M\'uzeum krt. 6--8, H-1088 Budapest,
Hungary, and\\
Research Institute
for Solid State Physics, P.O. Box 49, H--1525 Budapest, Hungary,}
\date{\today}
\maketitle
\vskip2pc]
\narrowtext

\begin{abstract}
The WKB-approximation for the Bogoliubov-equations of the
quasi-particle excitations in Bose-gases with condensate is worked out
in the case of spherically symmetric trap potentials
on the basis of the resulting quantization rule.
The excitation spectrum is calculated
numerically and also analytically in certain limiting cases.
%We show that there is a low-energy region, where the WKB results
%are in very good agreement with the results of quantum hydrodynamical
%calculations.
It is found that the energy levels of a Bohr-Sommerfeld type quantization
may be considerably shifted when the classical turning point
gets close to the surface of the condensate.
\end{abstract}


%\vskip2pc]
\narrowtext

The experimental realization and study of Bose-Einstein condensates
in alkali atom gases confined by magnetic traps
\cite{anderson,Li,Na,NaI,Rb}
has induced a vivid activity in the theoretical investigation
of such systems (See \cite{fetter_rev} for a recent review and
for further references). From a theoretical
point of view the existence of the external potential requires
new methods for calculating the physical properties of the
quantum gases with Bose-condensation.
Our aim is to solve the Bogoliubov-equations in WKB-approximation
and to determine the excitation
spectrum on the basis of the resulting quantization rule.

In the Bogoliubov-theory the
field operator can be expressed as a linear
combination of quasiparticle creation and annihilation operators.
The corresponding (nonuniform) expansion coefficients $u_j(\bbox{r})$
and $v_j(\bbox{r})$ obey the coupled linear Bogoliubov eigenvalue
equations \cite{fetter0}
\begin{equation}
\left( \begin{array}{cc}
\hat{H}_{HF} & -K(\bbox{r}) \\
-K^{*}(\bbox{r}) & \hat{H}_{HF} \\
\end{array} \right)
\left( \begin{array}{c}
u_j(\bbox{r}) \\ v_j(\bbox{r})
\end{array} \right)
=E_j
\left( \begin{array}{c}
u_j(\bbox{r}) \\ -v_j(\bbox{r})
\end{array} \right) \, ,
\label{eq:uvj}
\end{equation}
where $j$ denotes one of the quasiparticle states and $E_j$ is the
corresponding quasiparticle energy. The Hartree-Fock operator $\hat{H}_{HF}$
takes the form
\begin{equation}
\hat{H}_{HF}=-{\hbar^2 \over 2m}\Delta +U(\bbox{r})+ 2 |K(\bbox{r})| -\mu\, ,
\label{eq:hfop}
\end{equation}
where $U(\bbox{r})$ is the trap potential, $\mu$ is the chemical potential,
\begin{equation}
K(\bbox{r})={4 \pi \hbar^2 a \over m} \psi_0(\bbox{r})^2
\label{eq:Kr}
\end{equation}
denotes the potential-like contribution
of the condensate, whose wave-function $\psi_0(\bbox{r})$ is normalized as
$\int\, d^3 r |\psi_0(\bbox{r})|^2 =N_0$.
$N_0$ is the number of particles in the condensate and
$a$ is the $s$-wave scattering length. In the following we shall assume
that $a>0$.
The quasiparticle amplitudes $u_j(\bbox{r})$ and $v_j(\bbox{r})$ are
normalized according to \cite{fetter0}
\begin{equation}
\int d^3 r \left( u_j^*(\bbox{r}) u_k(\bbox{r}) -
                  v_j^*(\bbox{r}) v_k(\bbox{r}) \right) =\delta_{jk}\, .
\label{eq:uvnorm}
\end{equation}
For the sake of simplicity
we choose the external potential as spherically symmetric.
Moreover we shall take $\psi_0(\bbox{r})$ and hence also
$K(\bbox{r})$ as real and shall also make frequent use of
the Thomas-Fermi
approximation \cite{TFapp}, which leads to
\begin{equation}
|\psi_0(\bbox{r})|^2=\cases{ {m \over 4 \pi \hbar^2 a}
\left( \mu_{TF}-U(r) \right)
& if $r<r_{TF}$ \cr
0 & otherwise        \cr                }\, .
\label{eq:TFapp}
\end{equation}
Here,  $U(r_{TF})=\mu_{TF}$ and
$\mu_{TF}$ is fixed by normalization.

One can introduce
spherical coordinates $r$, $\theta$, $\phi$ and separate variables
in the usual way:
\begin{equation}
\left( \begin{array}{c} u_j(\bbox{r}) \\ v_j(\bbox{r}) \end{array} \right)
={1 \over r}
\left( \begin{array}{c} u_{nl}(r) \\ v_{nl}(r) \end{array} \right)
Y_{lm}(\theta,\phi) \, ,\nonumber
\end{equation}
where $j$ denotes the usual quantum numbers $(n,l,m)$ for
isotropic problems and the $Y_{lm}$ are the spherical harmonics.

To solve the coupled,
radial equations obtained from (\ref{eq:uvj}) it is advantageous to use
the linear combinations  \cite{griffin}
\begin{equation}
G_{nl}^{\pm}(r)= \left( u_{nl}(r) \pm v_{nl}(r) \right) \, ,
\label{eq:gpm}
\end{equation}
which satisfy the uncoupled equations
\begin{equation}
\left(\hat{H}_{HF}^2-K(r)^2-E^2 \mp[\hat{H}_{HF},K(r)]\right)G^{\pm}(r)=0\,.
\label{eq:g1}
\end{equation}
Here $[,]$ denotes the commutator.
(For brevity we have omitted the indices $n$ and $l$).
Furthermore, it follows from the original
equations that
\begin{equation}
G^{\pm}={1 \over E} \left( \hat{H}_{HF} \pm K(r) \right) G^{\mp} \, ,
\label{eq:gg}
\end{equation}
which is compatible with equations (\ref{eq:g1}).

Now the operator $\hat{H}_{HF}$ has the form
\begin{equation}
\hat{H}_{HF}=-{\hbar^2 \over 2m} {d^2 \over dr^2} +U_{eff}(r),
\label{eq:hfrad}
\end{equation}
where
\begin{equation}
U_{eff}(r)={\hbar^2 l(l+1) \over 2mr^2} +U(r)+2K(r)-\mu .
\end{equation}
In our WKB treatment we use Langer's rule by replacing
$l(l+1)$ by $(l+1/2)^2$. In the following $U_{eff}(r)$ is
considered as a classical potential.

We shall consider two types of solutions
\begin{eqnarray}
G^{+}(r) &=& e^{{i\over \hbar}\left(S_0+{\hbar \over i}S_1+
\ldots \right)}\, ,
\label{eq:cla}\\
G^{+}(r) &=& e^{-{1\over \hbar}\left(\tilde{S}_0+\hbar
\tilde{S}_1+ \ldots \right)} \, ,
\label{eq:clf}
\end{eqnarray}
with real functions $S_0(r),S_1(r),\ldots$ and
$\tilde{S}_0(r),\tilde{S}_1(r),\ldots$ respectively.
%in the classically forbiden region respectivelly.
Gathering terms having different powers of $\hbar$
one gets first order ordinary differental eqations for
the unknown quantities occuring in~(\ref{eq:cla}) and
(\ref{eq:clf}).

%%%%%%%%%%%%%%%%%%%%%%%%%%%
% Here follows the WKB-procedure
%%%%%%%%%%%%%%%%%%%%%%%%%%%
First we consider solutions of the form (\ref{eq:cla}). The $O(\hbar^0)$
equation
is the classical Hamilton-Jacobi equation for the radial action $S_0(r)$,
%\begin{equation}
%0=\left[{1 \over 2m}\left( {dS_0\over dr}\right)^2 +
%U_{eff}(r) \right]^2 -E^2-K^2(r) \, ,
%\end{equation}
from which one can express the classical radial momenta as
\begin{equation}  |p_r| \equiv \left| {dS_0 \over dr}\right|
= \sqrt{2m \bigl(\pm \sqrt{E^2+K^2} -U_{eff}\bigr)}
\label{eq:radmom}
\end{equation}

%In the following we shall use the Thomas-Fermi approximation.
%It then becomes apparent that only the plus sign is allowed
%under the square root in Eq.~(\ref{eq:radmom}).
%If one solves Eq.~(\ref{eq:radmom}) for
%the energy $E$ one obtains the classical one-particle
%Hamiltonian
%\begin{equation}
%E=H(p_r,r)=\sqrt{\left[ {p_r^2 \over 2m} +U_{eff}\right]^2-K^2(r)}\, .
%\label{eq:radham}
%\end{equation}
%This classical Hamiltonian occurs
%in several contexts: For a bulk condensate ($U(\bbox{r})=0$) this Hamiltonian
%gives the quasiparticle excitation-spectrum in the Bogoliubov-approximation,
%i.e. a linear phonon-spectrum for small wave-number excitations
%and a quadratic $k$ dependence for large $k$ values.
%Note that a similar Hamiltonian appears
%in the local density approximation \cite{lda}
%and in the calculation of phase fluctuations
%based on the Gross-Pitaevskii equation \cite{gior}.
%Using (\ref{eq:radham}) one obtains for the radial velocity \cite{cont}
We shall assume that $U_{eff}>0$ in which case only the plus sign is allowed  
to have $p_r$ real. This is the case
for instance in the Thomas-Fermi approximation (\ref{eq:TFapp}).
We introduce the radial velocity in the usual way
$v_r=\partial H /\partial p_r$ by regarding E in (\ref{eq:radmom}) as
the classical Hamiltonian $H(p_r,r)$. The obtained expression
\cite{cont}
\begin{equation}
v_r={\sqrt{E^2+K^2(r)}\over E}{p_r \over m}\, .
\label{eq:radvel}
\end{equation}
reflects the peculiarity of the classical quasi particle dynamics
in traps. The effective quasi-particle mass, which can be read off from
(\ref{eq:radmom}), is energy and
space-dependent. It approaches the particle mass at the boundary of the
condensate, but can become much smaller yet remains non-zero even
in the center of
very large condensates
in traps. This is a fundamental
difference to the untrapped case, where the limit
 $E \rightarrow 0$ can be taken, in which
the quasi-particle mass vanishes.

By solving (\ref{eq:g1}) with the ansatz (\ref{eq:cla}) up to $S_1$,
then using (\ref{eq:gg}) for $G^{-}(r)$, and finally transforming
back from $G^{\pm}(r)$ to $u_{nl}(r)$ and $v_{nl}(r)$,
particular solutions of the radial Bogoliubov equations are
obtained in the form
\begin{equation}
\left( {u_{nl}(r) \atop v_{nl}(r)} \right)\simeq Const \times
\left( {u_B(r) \atop v_B(r)} \right){1 \over \sqrt{|v_r|}}
e^{\pm {i \over \hbar}\int^r p_r (r)\, dr}\, ,
\label{eq:clasol}
\end{equation}
where
$ u^2_B =\left(\sqrt{1+(K(r)/E)^2}  + 1 \right)/2$,
$u_B^2-v_B^2=1$
are the generalizations of the usual Bogoliubov-coefficients
for the case without trapping potential.
Note that the classical probability distribution is
inversely proportional to $|v_r|$ as expected physically.

Solutions (\ref{eq:clasol}) are valid in the classically
allowed region, i.e., between the classical
turning points $r_{t1}$ and $r_{t2}>r_{t1}$ defined by the condition
$p_r(r_{ti})=0$, $i=1,2$. We shall assume that there are two
turning points only. There may be three cases. Case A: if
$r_{t1}<r_{TF}<r_{t2}$, in other words, the classical particle
enters the condensate, then leaves it, and returns back again etc.;
Case B: if $r_{TF}<r_{t1}<r_{t2}$, i.e., we have only a simple classical
motion in the trapping potential; Case C: if $r_{t1}<r_{t2}<r_{TF}$,
in which case the classical motion is confined to the condensate.

Next we construct solutions of the form (\ref{eq:clf})  proceeding similarly
as before. Using the ansatz (\ref{eq:clf})
in Eq.~(\ref{eq:g1}).
there can exist two different solutions for $\tilde{S}_0$,
\begin{eqnarray}
\left| q_r^{(i)} \right| &\equiv&
\left| {d \tilde{S}^{(i)}_0 \over dr}\right|=
\sqrt{2m \bigl(U_{eff}+(-1)^i \sqrt{E^2+K^2} \bigr)}, \nonumber \\
&&\quad i=1,2.
\label{eq:radmomi}
\end{eqnarray}
Both signs are allowed for example in the Thomas-Fermi approximation
(\ref{eq:TFapp}). The solution for $i=1$ is defined only outside
the classically accessible region, while the other one ($i=2$) is
permissible for all $r$-values, if $U_{eff}>0$ (as we suppose), and
represents a solution which can only occur in the two component quasi
particle dynamics.
Let us define  furthermore quantities $w_r^{(i)}$
similar as in (\ref{eq:radvel}) by the relations
\begin{equation}
w_r^{(i)}={\sqrt{E^2+K^2(r)}\over E}{q_r^{(i)} \over m}\, .
\end{equation}
For a smooth potential $U_{eff}(r)$ it can be shown
that normalizable eigenfunctions cannot contain
$\tilde{S}^{(2)}_0$ in a WKB solution.
However
for states whose radial wavelength near the characteristic radius 
of the condensate is large  
compared to the width of the boundary layer there,
%the situation is different if we use the Thomas-Fermi approximation
%for the condensate, because
the effective potential can no longer be treated as
smooth. Furthermore, such a type of contribution must always be
present in $v(r)$ asymptotically, if the condensate is restricted
to a finite region. 

Let us now consider the allowed solutions in case A.
%$r<r_{TF}$ ($i=1$) and $r>r_{TF}$ ($i=2$).
Requiring normalizibily and performing turning point matching
at $r_{tj}$ one obtains
\begin{eqnarray}
\left( {u_{nl} \atop v_{nl}}\right) &\simeq&
{C_{1j} \over \sqrt{|w_r^{(2)}|}}\left( {v_B \atop -u_B }\right)
\exp\left[{(-1)^j \over \hbar}\int_r^{r_{TF}} \!\! q_r^{(2)}(r)\, dr\right]
\nonumber \\
&& +{C_{2j} \over \sqrt{|Z_r|}} \left( {u_B \atop v_B }\right) F(r) ,
\label{eq:sr}
\end{eqnarray}
where  $j=1$ and $j=2$ correspond to $r<r_{TF}$ and $r>r_{TF}$
respectively.
$C_{1j}$ and $C_{2j}$ are arbitrary constants and
\begin{eqnarray}
F(r)&=&\exp\left[{(-1)^j \over \hbar}\int_r^{r_{tj}} q_r^{(1)}(r)\, dr\right],
\nonumber \\
Z_r&=&w_r^{(1)}, \qquad\mbox{for }
\cases{0<r<r_{t1}& $(j=1),$\cr r_{t2}<r& $(j=2),$\cr}
\end{eqnarray}
and
\begin{eqnarray}
F(r)&=&2 \sin\left[{(-1)^j \over \hbar}
\int_r^{r_{tj}} p_r(r)\, dr+{\pi \over 4}\right] ,
\nonumber\\
Z_r&=&v_r, \qquad\mbox{for }
\cases{r_{t1}<r<r_{TF}& $(j=1)$, \cr r_{TF}<r<r_{t2 } & $(j=2)$.\cr}
\end{eqnarray}
%The solution above can be transferred for $r>r_{TF}$ by
%replacing $r_{t1}$ with $r_{t2}$, introducing two new constants
%$C_3$ and $C_4$, interchanging the upper and
%lower bounds of the integrals, and restricting $r$ to the range
%of integration.

Requiring that $u(r)$, $v(r)$ and their first derivatives
are continuous at $r_{TF}$ one gets four homogeneous linear equations
for the four unknown constants. In order to get non-trival solutions
the determinant of the coefficient matrix
%(which is energy-dependent)
should vanish. This leads to the semiclassical quantization rule

\begin{eqnarray}
0&=&-{p_A \over \hbar}\cos\biggl( {I_A +I_B\over \hbar}\biggr)+
\sin \biggl( {I_A\over \hbar}+{\pi \over 4}\biggr)
 \sin \biggl( {I_B\over \hbar}+{\pi \over 4}\biggr)  \nonumber \\
&& \times \left[ {mL \over p_A^2}- \left({L \over 2E}\right)^2
 {\hbar p_B^2 \over 2 p_B^3+ \hbar m L}\right] \, ,
\label{eq:scq}
\end{eqnarray}
where we have introduced the notations $p_A \equiv p_r(r_{TF})$,
$p_B \equiv q_r^{(2)}(r_{TF})$,
$L=(\partial K /\partial r )_{r_{TF}+0}-
(\partial K /\partial r )_{r_{TF}-0}$,
$I_A=\int_{r_{t1}}^{r_{TF}} p_r \, dr$,
$I_B=\int_{r_{TF}}^{r_{t2}} p_r \, dr$.

%%%%%%%%%%%%%%%%%%%%%%%%%%%
% Here begins the Bohr quantization
%%%%%%%%%%%%%%%%%%%%%%%%%%%
Keeping only the first term on the right hand side of (\ref{eq:scq})
leads to the usual Bohr-Sommerfeld quantization rule
\begin{equation}
\left( n+{1 \over 2}\right) ={1 \over \pi\hbar}
\int_{r_{t1}}^{r_{t2}}\!\! dr \,
\sqrt{2m\left(\sqrt{E^2+K^2(r)}-U_{eff}(r)\right)}
\label{eq:bohrq}
\end{equation}
with the integer
radial quantum number $n \ge 0$ and including the Maslov-indices
due to the two turning
points in the radial motion.
The case B can be treated in an analogous way leading to the
quantization rule (\ref{eq:bohrq}) with $K(r)=0$ within
the range of integration. 

We discuss first the energy levels
on the basis of (\ref{eq:bohrq}) and will turn back to the
consequences of the general expression (\ref{eq:scq}) afterwards.
To evaluate (\ref{eq:bohrq}) we choose a harmonic potential
$U(r)=m \omega_0^2 r^2 /2$ often used in theoretical consideration
\cite{fetter_rev,TFapp,griffin,stringari}.
 To distinguish between
cases A and B let us use the dimensionless variables
$\tilde{J}=\hbar\omega_0(l+1/2)/(2\mu)$, $\tilde{E}=E/\mu$.
For energies and angular momenta in the region
$1<2\tilde{J}-1<\tilde{E}<\tilde{J}^2$
case B occurs
and the energy spectrum is simply that of a harmonic oscillator
shifted by $\mu$
\begin{equation}
E_{n,l}^{(osc)}=
\hbar \omega_0 \left( 2 n +l +{3 \over 2}\right) -\mu  .
\end{equation}
The self-consistency condition for
case B is then
$l+1/2>2\mu /\hbar \omega_0 +\sqrt{4\mu (2n+1)/\hbar \omega_0}$ .
One can check that case C,
i.e., the classical motion is entirely
inside the condensate, is not possible.

Considering the nontrivial case A in region
$0<\tilde{J}^2< \tilde{E}$
the action integral in Eq. (\ref{eq:bohrq})
can still be performed analytically (For the details see Ref. \cite{csord}),
but the result is rather cumbersome, and
the energies $E_{nl}$
%are given by implicit formulas, but
cannot be expressed explicitely. However, due to (\ref{eq:bohrq}) the
semiclassical energies fulfill the scaling relation
$E_{n,l}=\hbar \omega_0 G_{n,l}(\mu /\hbar \omega_0)$.
We discuss here some limiting cases. One interesting
limit is when one considers the high-lying levels, i.e., when
$E_{n,l} \gg \mu$ is fulfilled. Then the main contribution to the
action integral in (\ref{eq:bohrq}) comes from the region outside
of the condensate, leading to a spectrum which is  almost that of
a harmonic oscillator.
Expanding the action integral to the next to leading correction
in $\mu/E_{n,l}$ one gets:
\begin{eqnarray}
E_{n,l}&=&E_{n,l}^{(osc)}+\hbar \omega_0 \delta_{n,l} \nonumber \\
\delta_{n,l} &\simeq& {1 \over 3 \pi}
{\left[ {4 \mu \over \hbar \omega_0}
\left( 2 n +l +{3 \over 2} -{\mu \over \hbar \omega_0} \right)
-\left( l+{1 \over 2} \right)^2
\right]^{3/2}
\over
\left[ 2 n+l +{3 \over 2} -{\mu \over \hbar \omega_0} \right]^2}.
\end{eqnarray}
This result and that of the perturbation theoretical calculation
\cite{csord1} agree for large $(2 n + l)$ values.

The other interesting limit
is the  region of  excitation energies
small compared with the chemical potential. To reach it formally,
the angular momentum and the radial quantum numbers $l$ and $n$
are kept fixed but $\mu /\hbar \omega_0 $ tends to
infinity.
%In this limit  $E_{n,l}/\hbar \omega_0$ tends to
%a finite value as a function of $\mu /\hbar \omega_0 $.
The main contribution to the radial
action integral (\ref{eq:bohrq}) comes from those $r$ values,
which are within the condensate. % and the outside region is negligible.
%Performing this limit Eq. (\ref{eq:bohrq}) gives
To leading order:
\begin{equation}
E_{n,l} \simeq \hbar \omega_0
\left[ 2n^2 +2nl +3n +l +1 \right]^{1/2}\, .
\label{eq:ehydro}
\end{equation}
%This result clearly shows that in the large condensate limit
%the lower part of the excitation spectrum is independent of
%the chemical potential and of the number of the particles in the condensate
%($N_0$) and also independent of the $s$-wave scattering length.
Our result (\ref{eq:ehydro}) almost coincides
with that of  Stringari's hydrodynamic
calculation  \cite{stringari},
except for the last constant 1 within $[...]^{1/2}$ in
(\ref{eq:ehydro}), which has
an appreciable effect only on the lowest levels.
It is, of course, not unexpected that a WKB approach may fail there. For
somewhat higher energies at fixed but
large chemical potential there is a considerable region
where the two spectra calculated in WKB and in
hydrodynamical approximations, respectively, overlap.
For even higher energies the applicability of the
hydrodynamical approach looses its validity.
%%%%%%%%%%%%%%%%%%%%%%%%%%%
%  Numerics
%%%%%%%%%%%%%%%%%%%%%%%%%%%%
The task of solving (\ref{eq:bohrq}) for E can be carried out numerically
in a straightforward manner for given scattering length a,
trapping potential (i.e.
$\omega_0$)
and number $N_0$ of atoms
in the condensate, fixing the single parameter $2\mu/\hbar \omega_0=(15N_0
a/\sqrt{\hbar/m\omega_0})^{2/5}$
on which the spectrum depends.
An example of the results obtained is depicted in Figure~\ref{fig:ener}.
%Fixing the angular momentum quantum number $l$ the integral
%in (\ref{eq:bohrq}) is a monotonic function of $E$ because the
%classical radial frequency is positive. This fact helps to
%look for such an energy for which the right hand side is equal to
%a half-integer.
%
%In Figure~\ref{fig:comp} we compare our spectrum with that of
%Stringari. It is clearly seen that for fixed but large chemical
%potential our spectrum fails for small quantum numbers, which
%is not unexpected since
%the WKB-quantization gives  in general good
%results for high-lying levels, but has some defects for small
%energies. For somewhat higher  energies there is a large
%region where the two spectra overlap. For even higher energies
%the applicability of the hydrodynamical approximation fails.

%Let us briefly summarize our principal results:
%We have shown in this paper that
%the quasi-particles of a trapped
%Bose-condensate of weakly interacting atoms can
%be described in the classical limit by a particle-antiparticle dynamics
%with a
%square-root Hamiltonian in (15) reminiscent of the relativistic
%particle-antiparticle
%dynamics.
%We have then studied the semiclassical
%quantization of the particle-antiparticle
%dynamics by a procedure similar to the one employed for Dirac's equation.
% and obtained in analytical form an implicit quasi-classical expression
%for the energy levels
%which holds to an excellent approximation
%for all but the lowest levels and lends
%itself to straightforward numerical evaluation. We have also demonstrated how 
%known results in the literature are contained
%in our general result as limiting
%cases.

Let us turn now to the discussion of (\ref{eq:scq}). By solving
(\ref{eq:scq}) numerically for the energies we have found that the
corrections $\Delta E_{n,l}$ to the levels defined by (\ref{eq:bohrq})
are small except when the classical inner turning point $r_{t1}$ gets close
to the surface of the condensate (the border between regions A, B in the  
(n,l)-plane), in which case the radial wavenumber $p_A$ goes to zero. $\Delta  
E_{n,l}$ then becomes large but decreases rapidly
when going away from this situation. Note that in this case
the classical orbits are just glancing at the surface of the condensate.
The solution of the Bogoliubov equations then contains an anomalous
contribution, namely, the first term on the right hand side of
(\ref{eq:sr}), which is exponentially localized at the surface.
The effect remains even for high energies when the anomalous part of
$u$ and the usual part of $v$ become negligible.
At such energies it is generally assumed that the Bogoliubov equations
go over to the Hartree-Fock equations.
Our results suggest that there are exceptional states at the border of region  
A and B for which
this is not true due to the disturbance at the surface.
%We believe that the qualitative aspects of the effect may be
%present in a more complete treatment and are independent from
%the WKB and the Thomas-Fermi approximations, at least for such
%energies when the details of the structure of the surface
%are unimportant.
Physically,the effect is caused by the narrow boundary layer of the  
condensate which looks effectivly sharp for orbits glancing on the surface.  
Its qualitative aspects can therefore be expected to be independent from the  
WKB and Thomas-Fermi approximations. Experimentally such anomalous states  
could be observed by their excitation via modulations of the trapping  
potentials as
in \cite{NaI}, \cite{Rb} or by light scattering.
%%%%%%%%%%%%%%%%%%%%%%%%%%%%%%%%%%%%%%%%%%%%%%
%the Thomas-Fermi radius $r_{TF}$. Such an exceptional case
%($r_{t1}\approx r_{TF} \approx r_{t2}$) is realized
%for $n=0$ and $E<\mu$ when the classical periodic orbits run near the
%Thomas-Fermi surface, which leads to the absence of a solution $E_{0,l}$
%of (\ref{eq:scq}). Note that $n=0$ is anyhow outside the relevance of a
%WKB treatment. Furthermore for $E>\mu$ near the borderline of the case A and B
%$\Delta E_{n,l}$ can exceed $\hbar \omega_0$ due to the fact that
%$r_{t1}$ gets close to $r_{TF}$. ($\Delta E_{n,l}$ in case B
%can be derived analogously to Eq. (\ref{eq:scq})) However,
%the values of $\Delta E_{n,l}$ are much smaller generally in case B
%then in A, and in both cases decay fast when we go away from the
%critical line $E_{n,l}\approx (\hbar \omega_0)^2 (l+1/2)^2/4\mu$.
%Close to the critical line the WKB solution is not accurate because
%the vicinity of the Thomas-Fermi surface, where the potential
%is non-analytical, gets a large weight due to the large probability
%density near the classical turning points. For such energies a more
%accurate treatment is necessary.

In this paper we have restricted ourselves, for the sake of simplicity,
to the case of the spherically
symmetric trap potential.
Calculations along these lines for anisotropic harmonic oscillator
trap potentials as they are used in the experiments \cite{NaI,Rb} will
be published in a separate paper.
Here we only mention that the corresponding classical Hamiltonian
shows chaotic behavior \cite{fliesser}, especially for energies
comparable to the chemical potential.

We are indebted for useful discussions to A.~Voros and
G.~Vattay.
This work has been supported by the
project
of the Hungarian Academy of Sciences and the Deutsche
Forschungsgemeinschaft under grant No. 95.
%"Nonlinear dynamics of systems with Bose-Einstein condensate"
One of us (R.G.) wishes to acknowledge support by the Deutsche
Forschungsgemeinschaft through the Sonderforschungsbereich 237
``Unordnung und gro{\ss}e Fluktuationen''.
Two of us (A.Cs,P.Sz) would like to acknowledge support by
the Hungarian Academy of Sciences under grant No. AKP 96-12/12 and
by the Ministry of Education of Hungary under grant No. MKM 337.
The work has been partially supported by the Hungarian
National Scientific Research Foundation under grant
Nos. OTKA T017493 and F020094.

\begin{figure}
\caption{Energy levels $E(n,l)=E_{nl}/\hbar \omega_0$
obtained numerically from (\protect\ref{eq:bohrq})
%of Eq.~(\protect\ref{eq:uvj})
for an isotropic harmonic oscillator trap potential
$U(r)=m \omega_0^2 r^2/2$
as a function of the radial
quantum number $n$ and of the angular momentum quantum number $l$.
The chemical potential was chosen to be $8 \hbar\omega_0$
\label{fig:ener}}
\end{figure}


\begin{references}
\bibitem{cont} It is worth noting that the classical current related
to the velocity field has a quantum mechanical counterpart also in this
quasi-particle dynamics. To obtain it one has to subtitute for
the energy eigenvalue the time derivative operator
$i\hbar \partial/\partial t$ in (\protect\ref{eq:uvj})
in order to consider the time dependent
solutions $u(\bbox{r},t)$, $v(\bbox{r},t)$. Through the usual
steps one arrives at the continuity equation
$(\partial \rho / \partial t)+\mbox{div}\, \bbox{j}=0$, with
$\rho=|u|^2 - |v|^2$ and
$\bbox{j}=(\hbar /m)\mbox{Im}(u^*\nabla u+v^*\nabla v)$. The appearence
of different signs for the contributions of $v$ in the expressions
for $\rho$ and $\bbox{j}$ leads
to the unusual relation (\protect\ref{eq:radvel}) between velocity
and momentum in the classical limit.
\end{references}
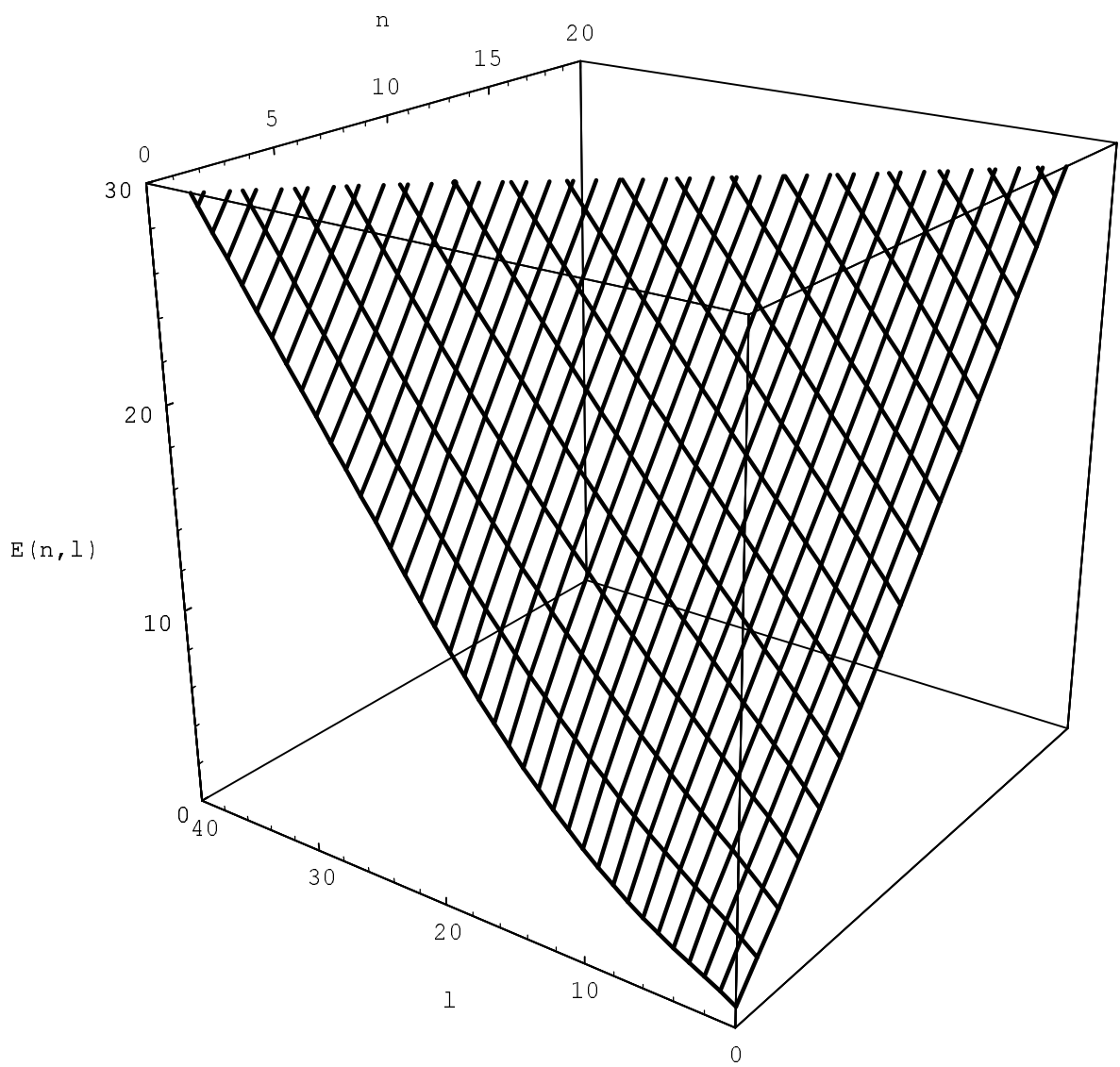
\end{document}